\documentclass{iaus}
\usepackage{graphicx}

\title[The VSOP Survey: Final aggregate results] 
{The VSOP Survey: Final aggregate results}

\author[Dodson \etal]   %% give here short author list %%
{Richard Dodson$^1$%
  \thanks{Present address: Marie Curie Fellow, OAN, Alcal\'a de Heneres, Espa\~na},
        S.~Horiuchi$^{2,3,4}$,
        W.~Scott$^{5}$,
        E.~Fomalont$^{6}$,
        Z.~Paragi$^{7}$,
        S.~Frey $^{8}$,
        K.~Wiik $^{1,9}$,
        H.~Hirabayashi$^{1}$,
        P.~Edwards$^{1,10}$,
        Y.~Murata$^{1}$,
        G.~Moellenbrock$^{11}$,
        L.~Gurvits$^{7}$,
        Z.~Shen$^{12}$,
        J.~Lovell$^{10}$
}

\affiliation{
$^{1}$The Institute of Space and Astronautical Science, JAXA,
                 3-1-1 Yoshinodai,
                 229-8510, Japan\\

$^{2}$Centre for Astrophysics and Supercomputing,
                  Swinburne University,
                  Vic. 3122, Australia\\

$^{3}$National Astronomical Observatory, 2-21-1 Osawa, 
                 Mitaka, Tokyo 181-8588, Japan\\

$^{4}$Jet Propulsion Laboratory, 4800 Oak Grove Drive,
                 Pasadena, CA 91109, USA\\

$^{5}$Physics and Astronomy Department, University of Calgary,
                 Calgary, Canada, T2N 1N4\\

$^{6}$National Radio Astronomy Observatory, 520 Edgemont Road,
                 Charlottesville, VA 22903, USA\\

$^{7}$Joint Institute for VLBI in Europe, P.O. Box 2,
                 7990 AA, Dwingeloo, Netherlands\\ 

$^{8}$F\"{O}MI Satellite Geodetic Observatory, P.O. Box 546,
                 H-1373, Budapest, Hungary\\

$^{9}$Tuorla Observatory,  
		 V\"ais\"al\"antie 20, FIN-21500 Piikki\"o, Finland\\

$^{10}$Australia Telescope National Facility,
                 CSIRO, 
                 P. O. Box 76, Epping NSW 2122, Australia\\

$^{11}$National Radio Astronomy Observatory, 
                 P.O. Box 0, Socorro, NM 87801, USA\\

$^{12}$Shanghai Astronomical Observatory, 
                  Chinese Academy of Sciences, 80 Nandan Lu, 
                  China
}

\pubyear{2006}
\volume{238}  %% insert here IAU Symposium No.
\pagerange{001--999}
\date{??? and in revised form ???}
\jname{Black Holes: from Stars to Galaxies -- across the Range of Masses}
\editors{V. Karas \& G. Matt, eds.}
\begin{document}

\maketitle

\begin{abstract}

In February 1997 the Japanese radio astronomy satellite HALCA was
launched to provide the space-borne element for the VSOP
mission. HALCA provided linear baselines three-times greater than that
of ground arrays, thus providing higher resolution and higher AGN
brightness temperature measurements and limits.  Twenty-five percent
of the scientific time of the mission was devoted to the ``VSOP
survey'' of bright, compact, extra-galactic radio sources at 5 GHz. A
complete list of 294 survey targets were selected from pre-launch
surveys, 91\% of which were observed during the satellite's lifetime.
The major goals of the VSOP Survey are statistical in nature: to
determine the brightness temperature and approximate structure, to
provide a source list for use with future space VLBI missions, and to
compare radio properties with other data throughout the
electro-magnetic spectrum.  All the data collected have now been
analysed and is being prepared for the final image Survey paper.  In
this paper we present details of the mission, and some statistics of
the images and brightness temperatures.


%% add here a maximum of 10 keywords, to be taken form the file <Keywords.txt>
\end{abstract}

\firstsection % if your document starts with a section,
              % remove some space above using this command.
\section{Introduction}

The {\bf V}LBI Space Observing Program (VSOP) satellite HALCA provided
the space baseline for the observation of a complete sample of bright
compact {\bf A}ctive Galactic Nuclei (AGNS) at 5~GHz; the VSOP survey
(Hirabayashi \etal\ 2000). Of this set of 294 AGNs: 102 were presented
in Scott \etal\ (2004) (hereafter P-III), 140 will be presented in
Dodson \etal\ (2007) (hereafter P-V), and 29 were not observed. The
remaining 23 did not produce space fringes, where we expected to find
them. Some of these may be correct, however we have erred on
the side of caution and consider them to be failures.
%Figure \ref{fig:hist}a) plots
%the contribution of all the antennae to the published, or to
%be published, experiments.
%, and Figure \ref{fig:hist}b) plots the final status of all the
%survey experiments.  

\begin{figure}
\begin{center}
 \includegraphics[width=0.9\textwidth]{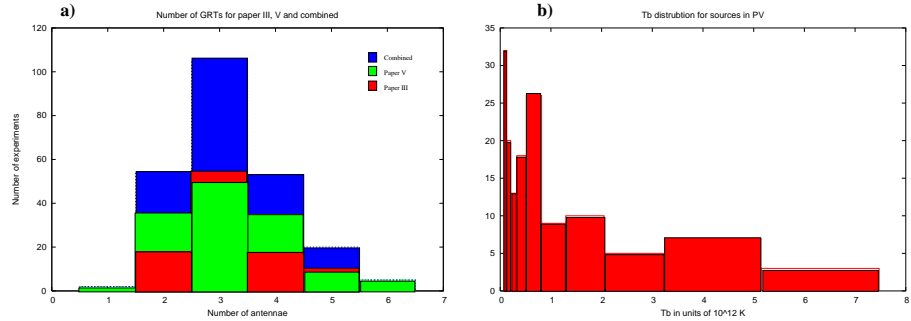}
  \caption{a) A histogram of the number of antennae for experiments
 in P-III, V and both. b) A histogram of the upper limits to the
 source rest frame brightness temperatures from the imaged data in
 P-V, in logarithmic bins.}\label{fig:hist}
\end{center}
\end{figure}

The HA{\bf L}CA downlink had a bandwidth of 128 Mbps, or 32 MHz at
Nyquist, two bit, sampling. A typical {\bf V}SOP Survey {\bf
e}xperiment would consist of three Ground Radio Telescopes (GRTs) and
a single tracking pass of HALCA; normally several hours. This leads to
a nomimal fringe sensitivity of 100mJy, and a image noise level
of 10mJy, for the space baselines. Figure \ref{fig:hist}a) shows the
number of antennae in each experiment. 
%%The images produced are compared, where possible, to both the General
%%Observing Time (GOT), and any other available VLBI, images. We
%%confirm, as reported in Lister \etal\ 2000, that the Survey data
%%reproduces that of the more completely sampled images.
%% %, see Figure \ref{fig:comb}
%% %which plots all the images from Dodson \etal (2007). 

One of the core goals of the VSOP Survey imaging program was to
measure lower limits to the brightness temperatures (T$_{\rm b}$)
directly from the data (c.f. Horiuchi \etal\ 2004). The determined
T$_{\rm b}$ (from VLBI) depends only on the physical baseline length,
independent of frequency, so space baselines will always provide the
highest possible limits. Figure \ref{fig:hist}b) plots the lower
limits to T$_{\rm b}$ for the sources in P-V, binned
logarithmically. These are the source frame T$_{\rm b}$ values, unless
the redshift is unknown. In which case the observer frame values are
used, as a lower limit. Lower limits were estimated from 
the lowest brightness temperature model compatible with the data. 

\section{Conclusions}\label{sec:concl}

We have completed the VSOP survey imaging data reduction. The paper
covering 140 sources is in preparation, and when it is published the
imaging portion of the VSOP Survey Project will be completed. 
We have directly measured the source bright{\bf n}ess temperatures,
and produce a distribution of the lower limits to directly measured
T$_{\rm b}$. 

%The VSOP project, as shown by the range of nationalities
%in this author list, required a deep international collaboration. The
%success of this bodes well for future projects, such as VSOP-2 and
%SKA. 

\begin{acknowledgments}

RD, KJW and J{\bf E}JL acknowledge support from the Japan Society for the
Promotion of Science.  WKS wishes to acknowledge support from the
Canadian Space Agency.  SH acknowledges support through an
N{\bf R}C/N{\bf A}SA-JPL Research Associateship. SF acknowledges the Bolyai
Scholarship received from the Hungarian Academy of Sciences. RD wishes
to thank S. Lorenzo. The NRAO is a facility of the National Science
Foundation, operated under cooperative agreement by Associated
Universities, Inc. The Australia Telescope Compact Array is part of
the Australia Telescope which is funded by the Commonwealth of
Australia for operation as a National Facility managed by CSIRO.

%% The contributions of very many people have been essential 
%% to the success of the VSOP Survey project. Many more than
%% can be included in the list of authors. We wish to recognise
%% and thank them for their contributions.

\end{acknowledgments}

%% \pagebreak
%% \begin{figure}
%%  \includegraphics[width=1.0\textwidth]{combined.eps}
%%   \caption{A montage of all the images which will appear in the
%%   VSOP Survey paper V.}\label{fig:comb}
%% \end{figure}

\end{document}